\documentclass{article}
\usepackage{ijcai22}

\usepackage{times}
\usepackage{soul}
\usepackage{url}
\usepackage[hidelinks]{hyperref}
\usepackage[utf8]{inputenc}
\usepackage[small]{caption}
\usepackage{graphicx}
\usepackage{amsmath}
\usepackage{amsthm}
\usepackage{booktabs}
\usepackage{algorithm}
\usepackage{algorithmic}
\usepackage{amssymb}

\usepackage{multirow}
\usepackage{xfrac}
\usepackage{xspace}
\usepackage{subcaption}
\newcommand{\ie}{\emph{i.e.},\xspace}
\newcommand{\eg}{\emph{e.g.},\xspace}

\def\B#1{\mathbf #1}
\def\C#1{\mathcal #1}

\title{Enhancing Sequential Recommendation with Graph Contrastive Learning}

\author{
Yixin Zhang$^{1}$\thanks{These two authors contributed equally to this work.}
\and
Yong Liu$^{3\ast}$\and
Yonghui Xu$^{2}$\thanks{Corresponding author}\and
Hao Xiong$^{5}$\and
Chenyi Lei$^{5}$\and
Wei He$^1$\and\\
Lizhen Cui$^{1,2\dag}$\And
Chunyan Miao$^{3,4}$\\
\affiliations
$^1$School of Software, Shandong University, China\\
$^2$Joint SDU-NTU Centre for Artificial Intelligence Research (C-FAIR), Shandong University, China\\
$^3$Alibaba-NTU Singapore JRI \& LILY Research Centre, Nanyang Technological University, Singapore\\
$^4$School of Computer Science and Engineering, Nanyang Technological University, Singapore\\
$^5$Alibaba Group, China\\
\emails
yixinzhang@mail.sdu.edu.cn, \{stephenliu, ascymiao\}@ntu.edu.sg, xu.yonghui@hotmail.com,\\
\{songling.xh, chenyi.lcy\}@alibaba-inc.com, \{hewei, clz\}@sdu.edu.cn
}

\begin{document}

\maketitle

\begin{abstract}

The sequential recommendation systems capture users' dynamic behavior patterns to predict their next interaction behaviors. Most existing sequential recommendation methods only exploit the local context information of an individual interaction sequence and learn model parameters solely based on the item prediction loss. Thus, they usually fail to learn appropriate sequence representations. This paper proposes a novel recommendation framework, namely Graph Contrastive Learning for Sequential Recommendation (GCL4SR). Specifically, GCL4SR employs a Weighted Item Transition Graph (WITG), built based on interaction sequences of all users, to provide global context information for each interaction and weaken the noise information in the sequence data. Moreover, GCL4SR uses subgraphs of WITG to augment the representation of each interaction sequence. Two auxiliary learning objectives have also been proposed to maximize the consistency between augmented representations induced by the same interaction sequence on WITG, and minimize the difference between the representations augmented by the global context on WITG and the local representation of the original sequence. Extensive experiments on real-world datasets demonstrate that GCL4SR consistently outperforms state-of-the-art sequential recommendation methods.

\end{abstract}

\section{Introduction}

In recent years, deep neural networks have been widely applied to build sequential recommendation systems~\cite{wang2019sequential,lei2021semi}. Although these methods usually achieve state-of-the-art sequential recommendation performance, there still exist some deficiencies that can be improved. \textit{Firstly}, existing methods model each user interaction sequence individually and only exploit the local context in each sequence. However, they usually ignore the correlation between users with similar behavior patterns (\eg with the same item subsequences).
\textit{Secondly}, the user behavior data is very sparse. Previous methods usually only use the item prediction task to train the recommendation models. They tend to suffer from the data sparsity problem and fail to learn appropriate sequence representations~\cite{zhou2020s3,xie2020contrastive}. \textit{Thirdly}, the sequential recommendation models are usually built based on implicit feedback sequences, which may include noise information (\eg accidentally clicking)~\cite{li2017neural}.

To remedy above issues, we first build a Weighted Item Transition Graph (WITG) to describe item transition patterns across the observed interaction sequences of all users. This transition graph can provide global context information for each user-item interaction~\cite{xu2019graph}. To alleviate the impacts of data sparsity, neighborhood sampling on WITG is performed to build augmented graph views for each interaction sequence. Then, graph contrastive learning~\cite{hassani2020contrastive,ZhangLZMWT22} is employed to learn augmented representations for the user interaction sequence, such that the global context information on WITG can be naturally incorporated into the augmented representations. Moreover, as WITG employs the transition frequency to describe the importance of each item transition, it can help weaken the impacts of noise interactions in the user interaction sequences, when learning sequence representations.

In this paper, we propose a novel recommendation model, named GCL4SR (\ie Graph Contrastive Learning for Sequential Recommendation). Specifically, GCL4SR leverages the subgraphs sampled from WITG to exploit the global context information across different sequences. The sequential recommendation task is improved by accommodating the global context information through the augmented views of the sequence on WITG. Moreover, we also develop two auxiliary learning objectives to maximize the consistency between augmented representations induced by the same interaction sequence on WITG, and minimize the difference between the representations augmented by the global context on WITG and the local representation of original sequence. Extensive experiments on public datasets demonstrate that GCL4SR consistently achieve better performance than state-of-the-art sequential recommendation approaches.

\section{Related Work}

In this section, we review the most relevant existing methods in sequential recommendation and self-supervised learning.

\subsection{Sequential Recommendation}

In the literature, Recurrent Neural Networks (RNN) are usually applied to build sequential recommendation systems. For example, GRU4Rec~\cite{hidasi2016session} treats users' behavior sequences as time series data and uses a multi-layer GRU structure to capture the sequential patterns. Moreover, some works, \eg NARM~\cite{li2017neural} and DREAM~\cite{feng2016dream}, combine attention mechanisms with GRU structures to learn users' dynamic representations. Simultaneously, Convolutional Neural Networks (CNN) have also been explored for sequential recommendation. Caser~\cite{tang2018personalized} is a representative method that uses both horizontal and vertical convolutional filters to extract users' sequential behavior patterns. Recently, SASRec~\cite{kang2018self} and BERT4Rec~\cite{sun2019bert4rec} only utilize self-attention mechanisms to model users' sequential behaviors. Beyond that, HGN~\cite{ma2019hierarchical} models users’ dynamic preferences using hierarchical gated networks. Along another line, Graph Neural Networks (GNN) have been explored to model complex item transition patterns. For instance, SR-GNN~\cite{wu2019session} converts sequences to graph structure data and employs the gated graph neural network to perform information propagation on the graph. GC-SAN~\cite{xu2019graph} dynamically builds a graph for each sequence and models the local dependencies and long-range dependencies between items by combining GNN and self-attention mechanism. In addition, GCE-GNN~\cite{wang2020global} builds the global graph and local graph to model global item transition patterns and local item transition patterns, respectively.

\subsection{Self-supervised Learning}

Self-supervised learning is an emerging unsupervised learning paradigm, which has been successfully applied in computer vision~\cite{jing2020self} and natural language processing~\cite{BERT2019}. There are several recent works applying self-supervised learning techniques in recommendation tasks. For example, \cite{zhou2020s3} maximizes the mutual information among attributes, items, and sequences by different self-supervised optimization objectives. \cite{xie2020contrastive} maximizes the agreement between two augmented views of the same interaction sequence through a contrastive learning objective. \cite{sigir21sgl} proposes a joint learning framework based on both the contrastive learning objective and recommendation objective. Moreover, in \cite{wei2021contrastive}, contrastive learning is used to solve the cold-start recommendation problem. In~\cite{ZhangLZMWT22}, a diffusion-based graph contrastive learning method is developed to improve the recommendation performance based on users' implicit feedback. Additionally, self-supervised learning has also been applied to exploit the item multimodal side information for recommendation~\cite{liu2021pre,lei2021understanding}.

\begin{figure}
\centering
    \includegraphics[width=0.95\columnwidth]{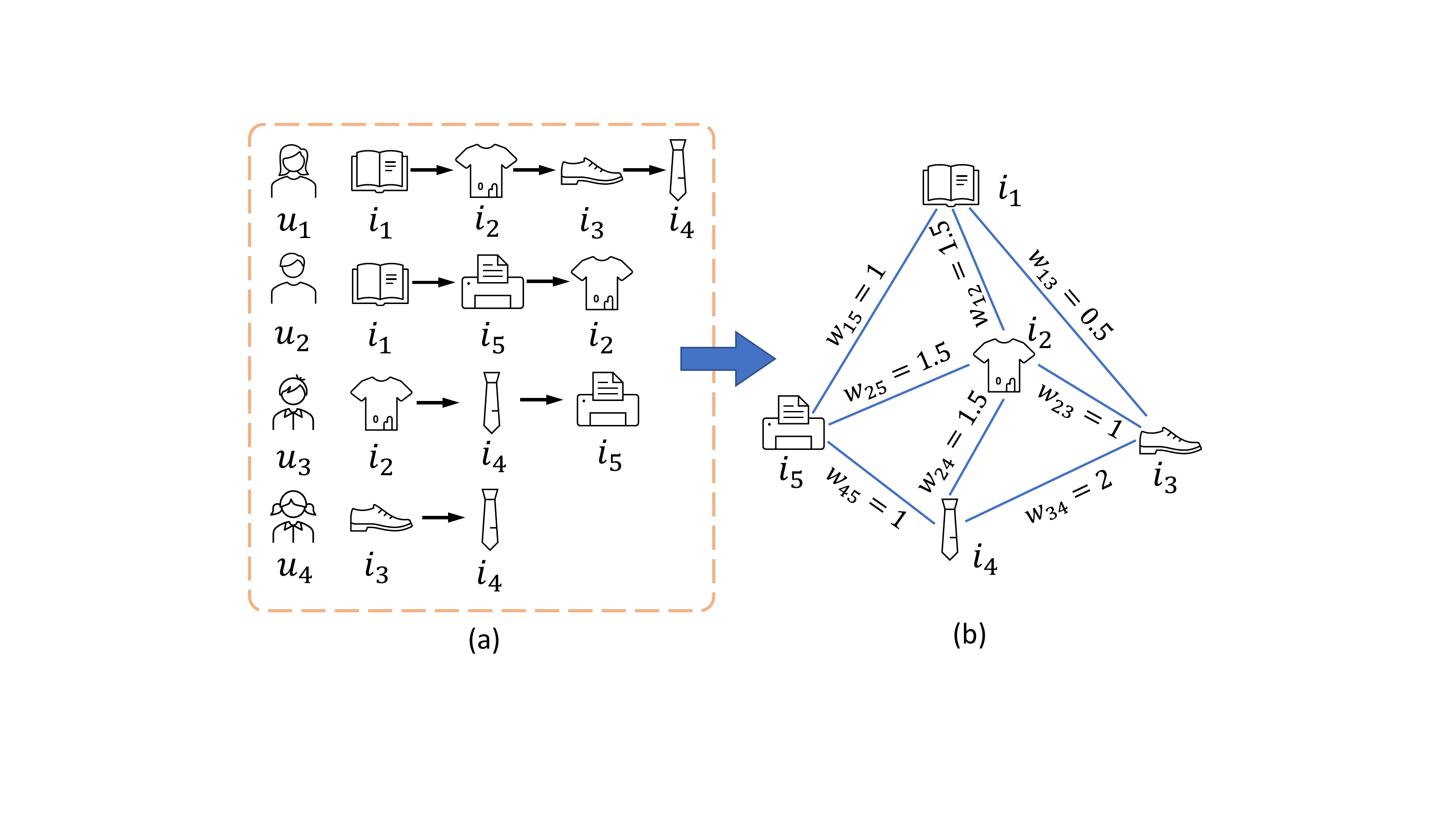}
    \caption{An example showing the transition graph construction procedure, where (a) shows the observed user behavior sequences, and (b) illuminates the weighted transition graph.}
    \label{fig:ttig}
\end{figure}

\begin{figure*}
\centering
    \includegraphics[width=0.85\textwidth]{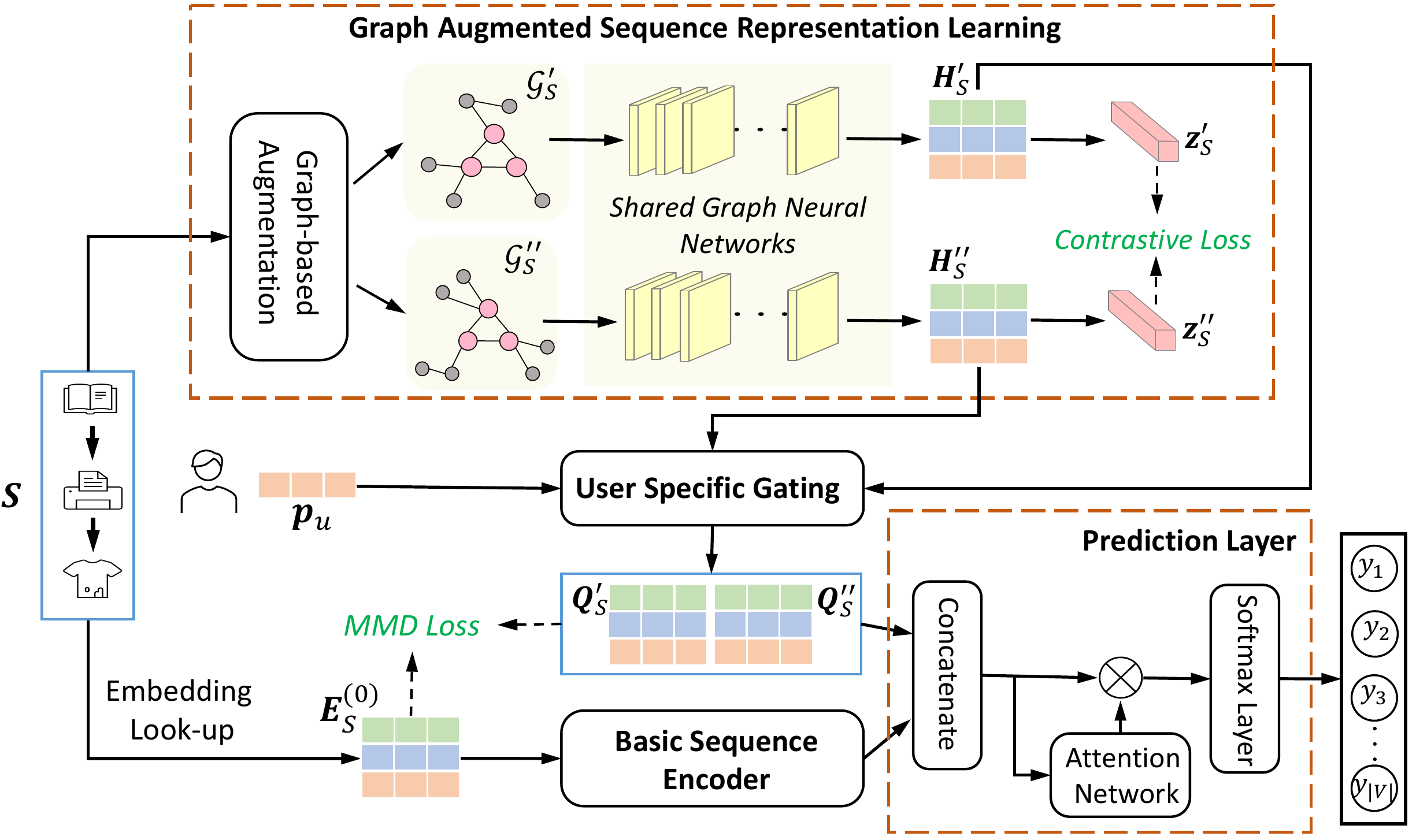}
    \caption{The framework of the proposed GCL4SR model.}
    \label{fig:framework}
\end{figure*}

\section{Preliminaries}

In this work, we study the sequential recommendation task, where we have the interaction sequences $\C{D}$ of a set of users $U$ over a set of items $V$. For each user $u \in U$, we use a list $S=\{v_1, v_2, \cdots, v_{n}\}$ to denote her interaction sequence, where $v_t$ is the $t$-th interaction item of $u$, and $n$ denotes the number of items that have interactions with $u$. Moreover, we denote the user $u$'s embedding by $\B{p}_u \in \mathbb{R}^{1\times d}$ and the item $i$'s embedding by $\B{e}_i \in \mathbb{R}^{1\times d}$. $\B{E}_S^{(0)} \in \mathbb{R}^{n\times d}$ is used to denote the initial embedding of the sequence $S$, where the $t$-th row in $\B{E}_S^{(0)}$ is the embedding of the $t$-th node in $S$. Similarly, $\B{E}\in \mathbb{R}^{|V| \times d}$ is used to denote the embeddings of all items.

Differing from existing methods that model the sequential transition patterns in each individual sequence,
we first build a weighted item transition graph $\C{G}$ from $\C{D}$ to provide a global view of the item transition patterns across all users' behavior sequences. The construction of the global transition graph $\C{G}$ follows the following strategy. Taking the sequence $S$ as an example, for each item $v_t \in S$, if there exists an edge between the items $v_t$ and $v_{(t+k)}$ in $\C{G}$, we update the edge weight as $w(v_t, v_{t+k}) \leftarrow w(v_t, v_{t+k}) + 1/k$; otherwise, we construct an edge between $v_t$ and $v_{t+k}$ in $\C{G}$ and empirically set the edge weight $w(v_t, v_{t+k})$ to $1/k$, where $k \in \{1, 2, 3\}$. Here, the score $1/k$ denotes the importance of a target node $v_t$ to its $k$-hop neighbor $v_{t+k}$ in the sequence. This empirical setting is inspired by the success of previous work~\cite{he2020lightgcn}. After repeating the above procedure for all the user sequences in $\C{D}$, we re-normalize the edge weight between two nodes $v_i$ and $v_j$ in $\C{G}$ as follows,
\begin{equation}\label{eq:weights}
  \widehat{w}(v_t, v_j) = w(v_t, v_j) \big(\frac{1}{\deg(v_i)} + \frac{1}{\deg(v_j)} \big),
\end{equation}
where $\deg(\cdot)$ denotes the degree of a node in $\C{G}$. Note that $\C{G}$ is an undirected graph. Figure~\ref{fig:ttig} shows an example about the transition graph without edge weight normalization.

\section{The Proposed Recommendation Model}

Figure~\ref{fig:framework} shows the overall framework of GCL4SR. Observe that GCL4SR has the following main components: 1) graph augmented sequence representation learning, 2) user-specific gating, 3) basic sequence encoder, and 4) prediction layer. Next, we introduce the details of each component.

\subsection{Graph Augmented Sequence Representation Learning}

\subsubsection{Graph-based Augmentation}

Given the weighted transition graph $\C{G}$, we first construct two augmented graph views for an interaction sequence $S$ through data augmentation. The motivation is to create comprehensively and realistically rational data via certain transformations on the original sequence.
In this work, we use the efficient neighborhood sampling method used in~\cite{hamilton2017inductive} to generate the augmented graph views from a large transition graph for a given sequence. Specifically, we treat each node $v \in S$ as a central node and interatively sample its neighbors in $\C{G}$ by empirically setting the sampling depth $M$ to 2 and the sampling size $N$ at each step to 20. In the sampling process, we uniformly sample nodes without considering the edge weights, and then preserve the edges between the sampled nodes and their weights in $\C{G}$. For a particular sequence $S$, after employing the graph-based augmentation, we can obtain two augmented graph views $\mathcal G_S^{'} = (V_S^{'}, E_S^{'}, \B{A}_S^{'})$ and $\mathcal G_S^{''} = (V_S^{''}, E_S^{''}, \B{A}_S^{''})$. Here, $V_S^{'}$, $E_S^{'}$, and $\B{A}_S^{'}$ are the set of nodes, the set of edges, and the adjacency matrix of $\C{G}_S^{'}$, respectively. Note that $\C{G}_S^{'}$ and $\C{G}_S^{''}$ are subgraphs of $\C{G}$, and the adjacency matrix $\B{A}_S^{'}$ and $\B{A}_S^{''}$ store the normalized weights of edges defined in Eq.~\eqref{eq:weights}.

\subsubsection{Shared Graph Neural Networks}

Following~\cite{hassani2020contrastive}, two graph neural networks with shared parameters are used to encode $\C{G}_S^{'}$ and $\C{G}_S^{''}$. Taking $\C{G}_S^{'}$ as an example, the information propagation and aggregation at the $t$-th layer of the graph neural networks are as follows,
\begin{align}
   \B{a}_{v_i}^{(t)} &= \mbox{Aggregate}^{(t)}\big(\{\B{h}_{v_j}^{(t-1)}: {v_j} \in N^{'}_{v_i}\}\big), \nonumber\\
   \B{h}_{v_i}^{(t)}&=\mbox{Combine}^{(t)}\big(\B{a}_{{v_i}}^{(t)}, \B{h}_{v_i}^{(t-1)}\big),
   \label{eq:GNN}
\end{align}
where $N^{'}_{v_i}$ denotes the set of $v_i$'s neighbors in $\C{G}_S^{'}$, $\B{h}_{v_i}^{(t-1)}$ denotes the representation of item $v_i$ at the $t$-th GNN layer. $\B{h}_{v_i}^{(0)}$ is $v_i$'s representation $\B{e}_i$ shared with the basic sequence encoder network.
In Eq.~\eqref{eq:GNN}, $\mbox{Aggregate}(\cdot)$ is the function aggregating the neighborhood information of a central node $v_i$, and $\mbox{Combine}(\cdot)$ is the function that combines neighborhood information to update the node embedding. After multiple layers of information propagation on $\C{G}_S^{'}$, we denote the embeddings of the nodes in $S$ at the last GNN layer by $\B{H}_S^{'} \in \mathbb{R}^{n \times d}$, which is the augmented representation of $S$ based on $\C{G}_S^{'}$. Similarly, we can obtain another augmented representation $\B{H}_S^{''} \in \mathbb{R}^{n \times d}$ of $S$ based on the augmented graph view $\C{G}_S^{''}$. In this work, these two GNNs are implemented as follows. At the first layer, we use Graph Neural Network (GCN) with the weighted adjacency matrix of an augmented graph to fuse the node information. Then, we further stack a GraphSage~\cite{hamilton2017inductive} layer that uses mean pooling to aggregate the high-order neighborhood information in the augmented graph.

\subsubsection{Graph Contrastive Learning Objective}

In this work, we use graph contrastive learning~\cite{hassani2020contrastive} to ensure the representations derived from augmented graph views of the same sequence to be similar, and the representations derived from augmented graph views of different sequences to be dissimilar. An auxiliary learning objective is developed to distinguish whether the two graph views are derived from the same user interaction sequence. Specifically, the views of the same sequence are used as positive pairs, \ie $\{(\C{G}_S^{'}, \C{G}_S^{''})|S\in \mathcal D\}$, and the views for different sequences are used as negative pairs, \ie $\{(\C{G}_S^{'}, \C{G}_K^{''})| S, K\in \mathcal D, S\neq K\}$. Then, we use the following contrastive objective to distinguish the augmented representations of the same interaction sequence from others,
\begin{equation}\label{eq8}
   \C{L}_{GCL}(S) = \sum_{S\in \mathcal D}{- \log\frac{\exp\big(cos(\B{z}_S^{'}, \B{z}_S^{''})/ \tau\big)}{\sum_{K\in \mathcal D} \exp\big(cos(\B{z}_S^{'}, \B{z}_K^{''})/\tau\big)}},
\end{equation}
where $\B{z}_S^{'}$ and $\B{z}_S^{''} \in \mathbb{R}^{1\times d}$ are obtained by performing mean pooling on $\B{H}_S^{'}$ and $\B{H}_S^{''}$ respectively, $cos(\cdot, \cdot)$ is the cosine similarity function, and $\tau$ is a hyper-parameter that is empirically set to 0.5 in the experiments.

\subsection{User-specific Gating}

As each individual user may only be interested in some specific properties of items, the global context information should be user-specific. Following~\cite{ma2019hierarchical}, we design the following user-specific gating mechanism to capture the global context information tailored to the user's personalized preferences,
\begin{align}
\B{Q}_S^{'}& = \B{H}_{S}^{'} \otimes  \sigma\big(\B{H}_{S}^{'} \B{W}_{g1} + \B{W}_{g2}\B{p}_u^\top),
\end{align}
where $\B{W}_{g1} \in {\mathbb{R}}^{d\times 1}$ and $\B{W}_{g2} \in {\mathbb{R}}^{L\times d}$, $\sigma(\cdot)$ is the sigmoid function, $\otimes$ is the element-wise product. Here, the user embedding $\B{p}_u$ describes the user's general preferences. Similarly, for augmented view $\C{G}_S^{''}$, we can obtain $\B{Q}_S^{''}$.

\subsubsection{Representation Alignment Objective}

The maximum mean discrepancy (MMD)~\cite{li2015generative} is then used to define the distance between the representations of personalized global context (\ie $\B{Q}_S^{'}$ and $\B{Q}_S^{''}$) and the local sequence representation $\B{E}_S^{(0)}$. Formally, the MMD between two feature distributions $\B{X}\in {\mathbb{R}}^{m \times d}$ and $\B{Y} \in {\mathbb{R}}^{\widetilde{m} \times d}$ can be defined as follows,
\begin{align}\label{eq11}
   & MMD(\B{X}, \B{Y}) = \frac{1}{m^2}\sum_{a=1}^m\sum_{b=1}^{m}{\mathcal K(\B{x}_a, \B{x}_b)} \nonumber\\
   & + \frac{1}{\widetilde{m}^2}\sum_{a=1}^{\widetilde{m}}\sum_{b=1}^{\widetilde{m}}{\mathcal K(\B{y}_a, \B{y}_b)} - \frac{2}{m\widetilde{m}}\sum_{a=1}^{m}\sum_{b=1}^{\widetilde{m}}{\mathcal K(\B{x}_a, \B{y}_b)},
\end{align}
where $\mathcal K(\cdot,\cdot)$ is the kernel function, $\B{x}_a$ and $\B{y}_b$ denote the $a$-th row of $\B{X}$ and the $b$-th row of $\B{Y}$, respectively. In this work, Gaussian kernel with bandwidth $\rho$ is used as the kernel function, \ie $\mathcal K(\B{x}, \B{x}^{'}) = e^{-\frac{||\B{x}-\B{x}^{'}||^2}{2\rho^2}}$.
Then, we minimize the distance between the representations of personalized global context and the local sequence representation as follows,
\begin{equation}\label{eq12}
   \C{L}_{MM}(S) = MMD\big(\B{E}_S^{(0)}, \B{Q}_S^{'}\big) + MMD\big(\B{E}_S^{(0)}, \B{Q}_S^{''}\big).
\end{equation}

\subsection{Basic Sequence Encoder}

Besides the graph augmented representations of a sequence, we also employ traditional sequential model to encode users' interaction sequences. Specifically, we choose SASRec~\cite{kang2018self} as the backbone model, which stacks the Transformer encoder~\cite{vaswani2017attention} to model the user interaction sequences. Given the node representation $\B{H}^{\ell-1}$ at the $(\ell-1)$-th layer, the output of Transformer encoder at the $\ell$-th layer is as follows,
\begin{align}\label{eq3}
    \B{H}^\ell &= \textit{FFN} \big(\textit{Concat}(head_1,\dots,head_h)\B{W}^h \big), \nonumber\\
   head_i &= \textit{Attention}\big({\B{H}^{\ell-1}\B{W}_i^Q},{\B{H}^{\ell-1}\B{W}_i^K},{\B{H}^{\ell-1}\B{W}_i^V}\big),
\end{align}
where $FFN(\cdot)$ denotes the feed-forward network, $h$ represents the number of heads, $\B{W}_i^Q, \B{W}_i^K, \B{W}_i^V \in \mathbb{R}^{d\times {\sfrac{d}{h}}}$, and $\B{W}^h\in {\mathbb{R}}^{d\times d}$ are the projection matrices. Specifically, we use the shared embedding $\B{E}_{S}^{(0)}$ with learnable position encoding as the initial state $\B{H}^0$. Here, the Residual Network, Dropout, and Layer Normalization strategies are omitted in the formula for convenience. Then, the attention mechanism is defined as,
\begin{equation}\label{eq2}
    \textit{Attention}(\B{Q}, \B{K}, \B{V})=\textit{softmax}\big(\frac{\B{Q}\B{K}^\top}{\sqrt{d}} \big)\B{V},
\end{equation}
where $\B{Q}$, $\B{K}$, and $\B{V}$ denote the queries, keys, and values respectively, and $\sqrt{d}$ is the scaling factor.

\subsection{Prediction Layer}

We concatenate the representations $\B{Q}_S^{'}$ and $\B{Q}_S^{''}$ obtained from the augmented graph views and the embeddings at the last layer of the Transformer encoder $\B{H}^\ell$ as follows,
\begin{equation}\label{eq13}
    \B{M} = \textit{AttNet}\big(\textit{Concat}\big(\B{Q}_S^{'}, \B{Q}_S^{''}, \B{H}^\ell \big)\B{W}_{T}\big),
\end{equation}
where $\B{M} \in \mathbb{R}^{1 \times d}$,  $\B{W}_{T} \in \mathbb{R}^{3d \times d}$ is the weight matrix, and $AttNet(\cdot)$ denotes the attention network. Then, given the user interaction sequence $S$ with length $n$, the interaction probabilities between the user and items at the $(n+1)$-th step can be defined as follows,
\begin{equation}
     \hat{\B{y}}^{(S)} = \emph{softmax}(\B{M} \B{E}^\top),
     \label{eq:prediction}
\end{equation}
where $\hat{\B{y}}^{(S)} \in \mathbb{R}^{1 \times |V|}$, and the $j$-th element of $\hat{\B{y}}^{(S)}$ denotes the interaction probability of the $j$-th item.

\subsection{Multi Task Learning}

Sequential recommendation aims to predict the next item that the user $u$ would like to interact with, based on her interaction sequence $S_u$ (Here, we include the subscript $u$ for clear discussion). Following~\cite{tang2018personalized},
we split the sequence $S_u=\{v_u^1, v_u^2, \cdots, v_u^{|S_u|}\}$ into a set of subsequences and target labels as follows: $\{(S_u^{1:1}, v_u^2), (S_u^{1:2}, v_u^3),\cdots,(S_u^{1:|S_u|-1}, v_u^{|S_u|})\}$, where $|S_u|$ denotes the length of $S_u$, $S_u^{1:k-1}=\{v_u^1, v_u^2, \cdots, v_u^{k-1}\}$, and $v_u^k$ is the target label of $S_u^{1:k-1}$.
Then, we formulate the following main learning objective based on cross-entropy,
\begin{equation}\label{eq:mainloss}
     \C{L}_{main} = -\sum_{S_u \in \C{D}}\sum_{k=1}^{|S_u|-1}\log\bigg( \hat{\B{y}}^{(S_u^{1:k})}(v_u^{k+1}) \bigg),
\end{equation}
where $\hat{\B{y}}^{(S_u^{1:k})}(v_u^{k+1})$ denotes the predicted interaction probability of $v_u^{k+1}$ based on the subsequence $S_u^{1:k}$ using Eq.~\eqref{eq:prediction}. In this work, we jointly optimize the main sequential prediction task and other two auxiliary learning objectives. The final objective function of GCL4SR is as follows, \begin{equation}\label{eq:finalobjective}
     \mathcal{L} = \mathcal{L}_{main} + \sum_{S_u \in \C{D}}\sum_{k=1}^{|S_u|-1}\lambda_{1} \mathcal{L}_{GCL}(S_u^{1:k}) + \lambda_{2} \mathcal{L}_{MM}(S_u^{1:k}),
\end{equation}
where $\lambda_1$ and $\lambda_2$ are hyper-parameters. The optimization problem in Eq.~\eqref{eq:finalobjective} is solved by a gradient descent algorithm.

\begin{table}[]
\centering
\small
\begin{tabular}{l|c|c|c|c}
\hline
& Home & Phones & Comics & Poetry \\\hline
\# Users & 66,519 & 27,879 & 13,810 & 3,522 \\\hline
\# Items & 28,237 & 10,429 & 16,630 & 2,624 \\ \hline
\# Interactions & 551,682 & 194,439 & 343,587 & 40,703 \\\hline
\# Nodes of $\C{G}$ & 28,237 & 10,429 & 16,630 & 2,624 \\ \hline
\# Edges of $\C{G}$ & 1,617,638 & 430,940 & 1,310,952 & 122,700 \\ \hline
\end{tabular}
\vspace{-5pt}
\caption{The statistics of experimental datasets.}
\vspace{-8pt}
\label{tab:datasets}
\end{table}

\begin{table*}[]
\centering
\small
\setlength\tabcolsep{0.8mm}{
\begin{tabular}{l|l|cccccccccccc}
\hline
Datasets                & Metrics &LightGCN & FPMC   & GRU4Rec & Caser  & SASRec & HGN    & SR-GNN & GC-SAN &GCE-GNN & CL4SRec & S$^3$-Rec  & GCL4SR \\ \hline
\multirow{4}{*}{Home}
                        & HR@10  & 0.0160  & 0.0162 & 0.0210  & 0.0101 & 0.0228 & 0.0152 & 0.0201 & \underline{0.0281} & 0.0259 & 0.0266  & 0.0280      & \textbf{0.0313}   \\
                        & HR@20  &0.0250 & 0.0218 & 0.0330  & 0.0173 & 0.0316 & 0.0231 & 0.0292 & 0.0394 & 0.0359 & 0.0387  & \underline{0.0406}      & \textbf{0.0422}   \\
                        & N@10 &0.0085 & 0.0097 & 0.0110  & 0.0051 & 0.0141 & 0.0083 & 0.0123 & \underline{0.0174} & 0.0161 & 0.0160  & 0.0169         & \textbf{0.0190}   \\
                        & N@20 &0.0108 & 0.0111 & 0.0140  & 0.0068 & 0.0163 & 0.0103 & 0.0146 & \underline{0.0197} & 0.0186 & 0.0186  & 0.0196         & \textbf{0.0218}   \\ \hline
\multirow{4}{*}{Phones}
                        & HR@10 &0.0687  & 0.0634 & 0.0835  & 0.0435 & 0.0883 & 0.0680 & 0.0778 & 0.0881 & 0.0946 & 0.0929  & \underline{0.1037}      & \textbf{0.1171}   \\
                        & HR@20 &0.1012  & 0.0854 & 0.1213  & 0.0647 & 0.1213 & 0.0990 & 0.1114 & 0.1232 & 0.1304 & 0.1305  & \underline{0.1428}      & \textbf{0.1666}   \\
                        & N@10 &0.0370 & 0.0374 & 0.0459  & 0.0233 & 0.0511 & 0.0364 & 0.0427 & 0.0500 & 0.0543 & 0.0533  & \underline{0.0594}         & \textbf{0.0665}   \\
                        & N@20 &0.0452 & 0.0430 & 0.0554  & 0.0287 & 0.0594 & 0.0442 & 0.0512 & 0.0588 & 0.0634 & 0.0627  & \underline{0.0693}         & \textbf{0.0790}   \\ \hline
\multirow{4}{*}{Poetry}
                        & HR@10 &0.1411  & 0.1275 & 0.1414  & 0.1068 & 0.1428 & 0.1034 & 0.1193 & 0.1309 & 0.1533 & 0.1496  & \underline{0.1613}      & \textbf{0.1638}   \\
                        & HR@20 &0.2127  & 0.1851 & 0.2104  & 0.1567 & 0.2030 & 0.1545 & 0.1723 & 0.1936 & 0.2229 & 0.2164  & \underline{0.2277}      & \textbf{0.2428}   \\
                        & N@10 &0.0771 & 0.0704 & 0.0783  & 0.0607 & 0.0829 & 0.0597 & 0.0686 & 0.0732 & 0.0859 & 0.0838  & \textbf{0.0915}            & \underline{0.0914}   \\
                        & N@20 &0.0954 & 0.0849 & 0.0956  & 0.0732 & 0.0980 & 0.0725 & 0.0818 & 0.0891 & 0.1035 & 0.1004  & \underline{0.1108}                     & \textbf{0.1112}   \\ \hline
\multirow{4}{*}{Comics}
                        & HR@10 &0.1106  & 0.1382 & 0.1593  & 0.1156 & 0.1709 & 0.1242 & 0.1481 & 0.1638 & 0.1722 & 0.1751  & \underline{0.1781} & \textbf{0.1829}   \\
                        & HR@20  &0.1672 & 0.1736 & 0.2058  & 0.1499 & 0.2100 & 0.1704 & 0.1857 & 0.2048 & 0.2232 & 0.2172  & \textbf{0.2258} & \underline{0.2249}   \\
                        & N@10 &0.0587 & 0.1019 & 0.1096  & 0.0790 & \underline{0.1276} & 0.0743 & 0.1067 & 0.1189 & 0.1222  & 0.1235  & 0.1234 & \textbf{0.1312}   \\
                        & N@20 &0.0730 & 0.1108 & 0.1213  & 0.0876 & \underline{0.1374} & 0.0859 & 0.1161 & 0.1292 & 0.1325 & 0.1341  & 0.1354 & \textbf{0.1417}   \\ \hline
\end{tabular}
}
\caption{The performance achieved by different methods. The best results are in \textbf{boldface}, and the second best results are \underline{underlined}.}
\label{table2}
\end{table*}

\section{Experiments}

In this section, we perform extensive experiments to evaluate the performance of the proposed GCL4SR method.

\subsection{Experimental Settings}

\paragraph{Datasets.} The experiments are conducted on the Amazon review dataset~\cite{he2016ups} and Goodreads review dataset~\cite{wan2019fine}. For Amazon dataset, we use two 5-core subsets for experimental evaluations: ``Home and Kitchen'' and ``Cell Phones and Accessories" (respectively denoted by Home and Phones). For Goodreads dataset, we choose users' rating behaviors in ``Poetry" and ``Comics Graphic" categories for evaluation.
Following~\cite{zhou2020s3}, we treat each rating as an implicit feedback record. For each user, we then remove duplicated interactions and sort her historical items by the interaction timestamp chronologically to obtain the user interaction sequence. To guarantee each user/item has enough interactions, we only keep the ``5-core" subset of each dataset, by iteratively removing the users and items that have less than 5 interaction records. Table~\ref{tab:datasets} summarizes the statistics of experimental datasets.

\paragraph{Setup and Metrics.} For each user, the last interaction item in her interaction sequence is used as testing data, and the second last item is used as validation data. The remaining items are used as training data. This setting has been widely used in previous studies~\cite{kang2018self,sun2019bert4rec,zhou2020s3}. The performance of different methods is assessed by two widely used evaluation metrics: Hit Ratio@$K$ and Normalized Discounted Cumulative Gain@$K$ (respectively denoted by HR@$K$ and N@$K$), where $K$ is empirically set to 10 and 20. For each metric, we first compute the accuracy for each user on the testing data, and then report the averaged accuracy for all testing users. In the experiments, all the evaluation metrics are computed on the whole candidate item set without negative sampling.

\paragraph{Baseline Methods.} We compare GCL4SR with the following baseline methods.
\begin{itemize}
  \item \textbf{LightGCN}~\cite{he2020lightgcn}: This method is a graph-based collaborative filtering light convolution network.

  \item \textbf{FPMC}~\cite{rendle2010factorizing}: This method combines matrix factorization and Markov chain model for sequential recommendation.

  \item \textbf{GRU4Rec}~\cite{hidasi2016session}: This method employs Gated Recurrent Unit (GRU) to capture the sequential dependencies and make recommendation.

  \item \textbf{Caser}~\cite{tang2018personalized}: This method uses both vertical and horizontal convolution to capture users’ sequential behavior patterns for recommendation.

 \item \textbf{SASRec}~\cite{kang2018self}: It uses self-attention mechanism to capture users' sequential patterns for recommendation.

 \item \textbf{HGN}~\cite{ma2019hierarchical}: This method uses a hierarchical gating network with an item-item product module for the sequential recommendation.

 \item \textbf{SR-GNN}~\cite{wu2019session}: This method converts sequences into graphs and leverages gated GNN layer to capture the item dependencies.

 \item \textbf{GC-SAN}~\cite{xu2019graph}: This method utilizes graph neural network and self-attention mechanism to dynamically capture rich local dependencies.

 \item \textbf{GCE-GNN}~\cite{wang2020global}: This method proposes to build global graph and local graph to model global transition patterns and local transition patterns, respectively.

 \item \textbf{S$^3$-Rec}~\cite{zhou2020s3}: This method employs different self-supervised optimization objectives to maximize the mutual information among attributes, items, and sequences.

  \item \textbf{CL4SRec}~\cite{xie2020contrastive}: This method uses sequence-level augmentation to learn better sequence representations.

\end{itemize}

\paragraph{Implementation Details.} All the evaluation methods are implemented by PyTorch~\cite{paszke2019pytorch}. Following~\cite{zhou2020s3}, we set the maximum sequence length to 50. The hyper-parameters of baseline methods are selected following the original papers, and the optimal settings are chosen based on the model performance on validation data. In the evaluated methods, only S$^3$-Rec considers the item attribute information, and the other methods do not use item attribute information. For fair comparison, we only keep the masked item prediction and segment prediction tasks of S$^3$-Rec to learn the model. For CGL4SR, we empirically set the number of self-attention blocks and attention heads to 2. The dimensionality of embeddings is set to 64. The weights for the two self-supervised losses $\lambda_{1}$ and $\lambda_{2}$ are chosen from $\{0.01, 0.05, 0.1, 0.3, 0.5, 0.7, 1.0\}$. We use Adam~\cite{kingma2014adam} as the optimizer and set the learning rate, $\beta_{1}$, and $\beta_{2}$ to 0.001, 0.9, and 0.999 respectively. Step decay of the learning rate is also adopted. The batch size is chosen from \{256, 512, 1024\}. The $L_2$ regularization coefficient is set to $5\times 10^{-5}$. We train the model with early stopping strategy based on the performance on validation data.

\subsection{Performance Comparison}

The performance comparison results are summarized in Table~\ref{table2}. Overall, GCL4SR outperforms all baseline methods on all datasets, in terms of almost all evaluation metrics.

Compared with RNN and CNN based models (\eg GRU4Rec and Caser), the models based on self-attention mechanism (\eg SASRec, GC-SAN, and GCL4SR) usually achieve better performance. This is because that self-attention mechanism is more effective in capturing long-range item dependencies. GC-SAN achieves better performance than SASRec on Home dataset, by introducing a graph built based on an individual sequence to improve the sequence representation. GCE-GNN outperforms SR-GNN by additionally exploiting global-level item transition patterns.

Moreover, CL4SRec, S$^3$-Rec, and GCL4SR usually outperform their backbone structure SASRec. This demonstrates that the self-supervised learning objectives can help improve sequential recommendation performance. In addition, GCL4SR achieves better results than CL4SRec and S$^3$-Rec. This is because that CL4SRec and S$^3$-Rec augment the sequence representation by the auxiliary learning objectives that only exploit the local context in each individual sequence. However, GCL4SR augments the sequence representation using subgraphs of the transition graph built based on sequences of all users, which can provide both local and global context for learning sequence representations.

\begin{figure*}[t!]
     \centering
     \begin{subfigure}[]{0.31\textwidth}
         \centering
         \includegraphics[width=\textwidth]{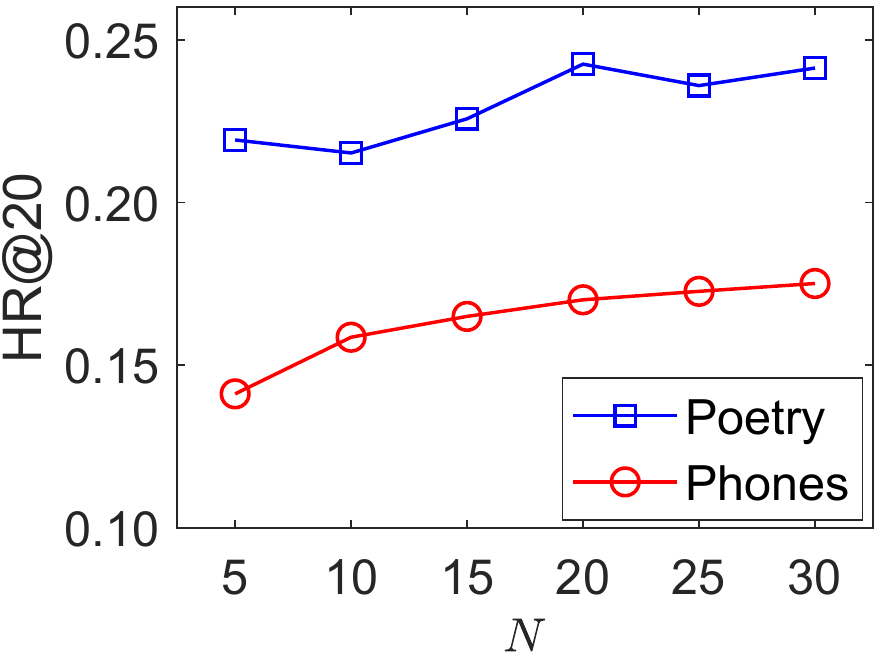}
         \caption{}
         \label{fig:H}
     \end{subfigure}\hfill
     \begin{subfigure}[]{0.31\textwidth}
         \centering
         \includegraphics[width=\textwidth]{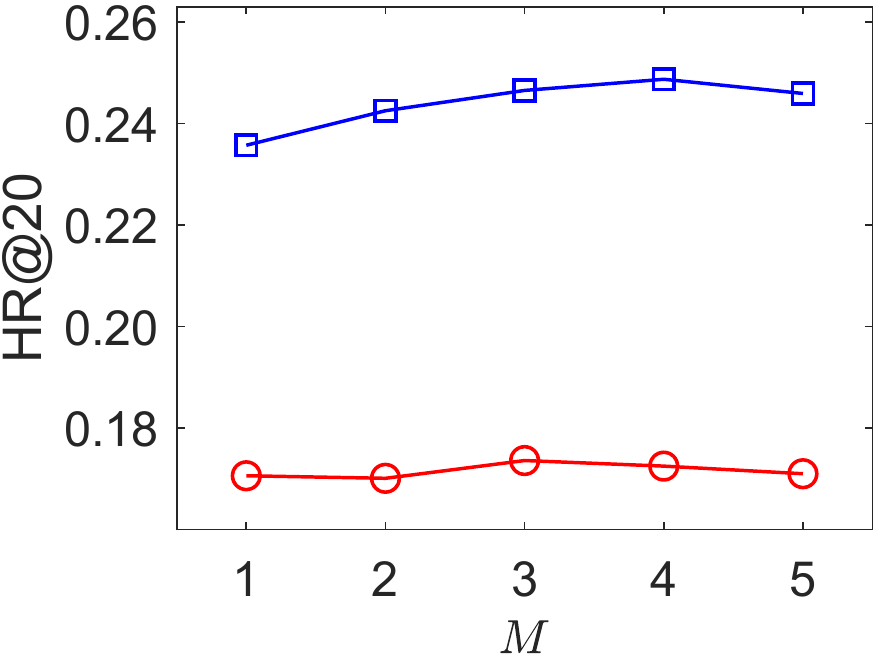}
         \caption{}
         \label{fig:L}
     \end{subfigure}\hfill
     \begin{subfigure}[]{0.31\textwidth}
         \centering
         \includegraphics[width=\textwidth]{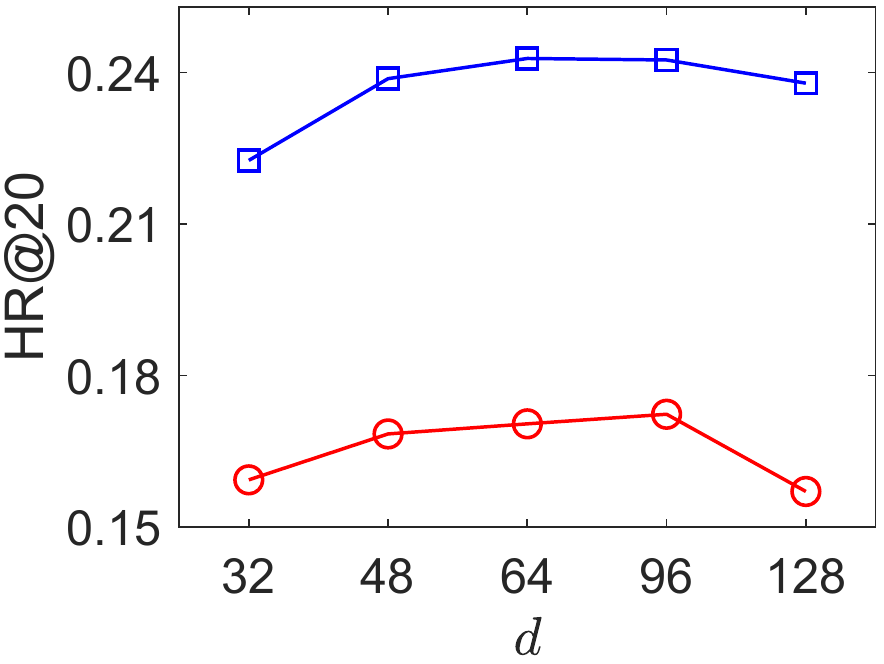}
         \caption{}
         \label{fig:D}
     \end{subfigure}
     \caption{The performance trends of GCL4SR with respect to different settings of $M$, $N$, and $d$ on Poetry and Phones datasets.}
     \label{fig:parameters}
\end{figure*}

\subsection{Ablation Study}

To study the importance of each component of GCL4SR, we consider the following GCL4SR variants for evaluation: 1) \textbf{GCL4SR}\textsubscript{w/o G}: we remove the graph contrastive learning loss by setting $\lambda_1$ to 0 in Eq.~\eqref{eq:finalobjective}; 2) \textbf{GCL4SR}\textsubscript{w/o GM}: we remove both the graph contrastive learning loss and the MMD loss by setting $\lambda_1$ and $\lambda_2$ to 0 in Eq.~\eqref{eq:finalobjective}; 3) \textbf{GCL4SR}\textsubscript{w/o W}: we remove the edge weights of the augmented graph views when performing GCN operations at the first layer of the shared GNNs used in graph contrastive learning.

Table~\ref{tab:ablation} summarizes the performance of GCL4SR variants and SASRec on Poetry and Phones datasets. We can note that GCL4SR\textsubscript{w/o GM} outperforms the backbone model SASRec in terms of HR@20, on both datasets. This indicates the context in the global transition graph can help improve sequential recommendation performance. By including the MMD loss, GCL4SR\textsubscript{w/o G} achieves better performance than GCL4SR\textsubscript{w/o GM}. By further combining the graph contrastive learning loss, GCL4SR outperforms GCL4SR\textsubscript{w/o G}, in terms of N@20, on both datasets. These observations demonstrate that both the MMD loss and graph contrastive learning loss can help learn better item and sequence representations for sequential recommendation. Moreover, GCL4SR outperforms GCL4SR\textsubscript{w/o W} in terms of all metrics. This observation indicates that the transition frequency between items across all sequences can help distinguish the importance of neighboring items for better sequential recommendation performance.

\begin{table}[]
\centering
\small
\begin{tabular}{l|cc|cc}
\hline
\multirow{2}{*}{Method} & \multicolumn{2}{c|}{Poetry}                              & \multicolumn{2}{c}{Phones}                              \\ \cline{2-5}
        & HR@20   & N@20 & HR@20   & N@20 \\ \hline
GCL4SR  & 0.2428 & \textbf{0.1112}    & \textbf{0.1666}  & \textbf{0.0790} \\
GCL4SR\textsubscript{w/o G} & \textbf{0.2433}  & 0.1095   & 0.1607  & 0.0734  \\
GCL4SR\textsubscript{w/o GM} & 0.2138  & 0.0958  & 0.1423 & 0.0713 \\
GCL4SR\textsubscript{w/o W} & 0.2172  & 0.0979   & 0.1500  & 0.0694  \\\hline
SASRec  & 0.2030  & 0.0980  & 0.1213  & 0.0594  \\\hline
\end{tabular}
\caption{The performance achieved by GCL4SR variants and SASRec on Poetry and Phones datasets.}
\label{tab:ablation}
\end{table}

\subsection{Parameter Sensitivity Study}

We also perform experiments to study the impacts of three hyper-parameters: the sampling depth $M$ and sampling size $N$ used in graph-based augmentation, and the embedding dimension $d$. Figure~\ref{fig:parameters} shows the performance of GCL4SR with respect to different settings of $M$, $N$, and $d$ on Poetry and Phones datasets. As shown in Figure~\ref{fig:parameters}(a), larger sampling size tends to produce better recommendation performance. For the sampling depth, we can notice the best settings for $M$ are 4 and 3 on Poetry and Phones datasets, respectively. In addition, the best performance is achieved by setting $d$ to 64 and 96 on Poetry and Phones datasets, respectively.

\begin{table}[]
\centering
\small
\begin{tabular}{l|cc|cc}
\hline
\multirow{2}{*}{Method} & \multicolumn{2}{c|}{Poetry} & \multicolumn{2}{c}{Phones}\\ \cline{2-5}
        & HR@20   & N@20 & HR@20   & N@20 \\ \hline
HGN  & 0.1545 & 0.0725    & 0.0990  & 0.0442 \\
GCL4SR-HGN & 0.1712  &  0.0763   & 0.1064  & 0.0475 \\\hline
GRU4Rec  & 0.2104 & 0.0956 & 0.1213 & 0.0554 \\
GCL4SR-GRU & 0.2362  & 0.1057  & 0.1622 & 0.0763 \\\hline
SASRec  & 0.2030  & 0.0980  & 0.1213  & 0.0594  \\
GCL4SR-SAS  & \textbf{0.2428} & \textbf{0.1112}    & \textbf{0.1666}  & \textbf{0.0790} \\
\hline
\end{tabular}
\caption{The performance of HGN, GRU4Rec, SASRec, and GCL4SR with different basic sequence encoders.}
\label{tab:encoder}
\end{table}

\subsection{Impacts of Sequence Encoders}

To further investigate the effectiveness of the graph augmented sequence representation learning module, we employ other structures to build the basic sequence encoder. Specifically, we consider the following settings of GCL4SR for experiments: 1) \textbf{GCL4SR-GRU}: we use the GRU4Rec as the backbone structure to build the basic sequence encoder; 2) \textbf{GCL4SR-HGN}: we use HGN as the backbone structure to build the basic sequence encoder; 3) \textbf{GCL4SR-SAS}: The default model that uses SASRec as the backbone structure to build the sequence encoder.

Table~\ref{tab:encoder} shows the performance of GCL4SR with different sequence encoders, as well as the performance of backbone models. Observe that GCL4SR-HGN, GCL4SR-GRU, and GCL4SR-SAS outperform the corresponding backbone encoder models. This indicates that the graph augmented sequence representation learning module is a general module that can help improve the performance of existing sequential recommendation methods. Moreover, GRU4Rec and SASRec achieve better performance than GCL4SR-HGN. This indicates the basic sequence encoder dominates the performance of GCL4SR, and the graph augmented sequence representation learning module is a complementary part that can help further improve the recommendation performance.

\section{Conclusion and Future Work}

This paper proposes a novel recommendation model, namely GCL4SR, which employs a global transition graph to describe item transition patterns across the interaction sequences of different users. Moreover, GCL4SR leverages the subgraphs randomly sampled from the transition graph to augment an interaction sequence. Two auxiliary learning objectives have been proposed to learn better item and sequence representations. Extensive results on real datasets demonstrate that the proposed GCL4SR model consistently outperforms existing sequential recommendation methods. For future work, we would like to develop novel auxiliary learning objectives to improve the performance of GCL4SR. Moreover, we are also interested in applying GCL4SR to improve the performance of other sequential recommendation models.

\section*{Acknowledgments}

This work is supported, in part, by the NSFC No.91846205, National Key R\&D Program of China No.2021YFF0900800, SDNSFC No.ZR2019LZH008, Shandong Provincial Key Research and Development Program (Major Scientific and Technological Innovation Project) (NO.2021CXGC010108), the Fundamental Research Funds of Shandong University. This work is also supported, in part, by Alibaba Group through Alibaba Innovative Research (AIR) Program and Alibaba-NTU Singapore Joint Research Institute (JRI), Nanyang Technological University, Singapore.

\bibliographystyle{named}
\bibliography{ijcai22}

\end{document}